\documentclass[11pt]{article}
\pdfoutput=1

\usepackage[T1]{fontenc}
\usepackage{fancyhdr}
\usepackage{amsmath}    
\usepackage{amssymb}
\usepackage{amsthm}
\usepackage{epsfig}
\usepackage{graphicx}
\usepackage{booktabs}

\newcommand{\RR}{\mathbb{R}}

\newtheorem{thm}{Theorem}
\newtheorem{defn}{Definition}

\def\doublespace{\baselineskip=\normalbaselineskip \multiply\baselineskip by 7
\divide\baselineskip by 5}

\begin{document}

\doublespace

\title{Nonparametric Partial Importance Sampling for Financial Derivative Pricing} 
\author{Jan C.\ Neddermeyer\footnote{University of Heidelberg,
Institute of Applied Mathematics, Im Neuenheimer Feld
294, D-69120 Heidelberg, Germany (e-mail: jc@neddermeyer.net). This work has
been supported by the Deutsche Forschungsgemeinschaft (DA 187/15-1).}\\\\
}
\date{}

\begin{titlepage}
\maketitle
\thispagestyle{empty}

\noindent
\textbf{Abstract.}
Importance sampling is a promising variance reduction technique
for Monte Carlo simulation based derivative pricing. Existing importance
sampling methods are based on a parametric choice of the proposal.
This article proposes an algorithm that estimates the
optimal proposal nonparametrically using a multivariate
frequency polygon estimator. In contrast to parametric methods, nonparametric
estimation allows for close approximation of the optimal proposal.
Standard nonparametric importance sampling is inefficient for high-dimensional
problems. We solve this issue by applying the procedure to a low-dimensional
subspace, which is identified through principal component analysis and the
concept of the effective dimension. The mean square error properties of the
algorithm are investigated and its asymptotic optimality is shown.
Quasi-Monte Carlo is used for further improvement of the method.
It is easy to implement, particularly it does not require any analytical
computation, and it is computationally very efficient.
We demonstrate through path-dependent and multi-asset option pricing problems
that the algorithm leads to significant efficiency gains compared to other
algorithms in the literature.


\end{titlepage}

\setcounter{page}{1}

\vspace{0.5cm}
\begin{center}
1. INTRODUCTION
\end{center}
In the last decade, the complexity of the pricing models used for 
evaluation of financial products experienced a
distinct increase. 
As a consequence of this development, pure numerical methods became more and
more inadequate for the arisen high-dimensional integration tasks. Often, Monte
Carlo (MC) integration is the only feasible method. This stems from the fact
that the MC convergence rate is independent of the problem dimension. 
However, crude MC is often inefficient for practical sample sizes.
Raising computing power and increasing the sample size is no solution. 
The need of efficient MC procedure is apparent.

To make MC algorithms comparable, it is useful to quantify
the efficiency of an estimator.
Let's assume $X$ is a random variable defined on a probability space
$(\Omega, \mathcal{B}, \mathbf{P})$ and it is used to estimate some quantity
$\mu$. The computational efficiency (CE) of estimator $X$ can be defined through
\begin{equation*}
\text{CE}[X] = (\text{MSE}[X] C[X])^{-1},
\end{equation*}
where $\text{MSE}[X]$ denotes the mean square error of $X$ and $C[X]$ the
average costs of computing one realization of $X$ (L'Ecuyer 1994). From this
definition one observes that efficiency improvements can be achieved either by 
reducing the MSE or the computational costs. 
The former includes well-known variance reduction (VR)
techniques such as importance sampling (IS), antithetic sampling, moment
matching, and control variates (J{\"a}ckel 2002; Glasserman
2004; Robert and Casella 2004). Most VR techniques aim at
improving a given set of samples, that is used for MC integration.
In contrast, IS is based on changing the distribution from which the samples are
drawn. The idea behind IS is to select a distribution (which is known as
proposal) that forces the samples into the
domain which is most important to the integrand.
Intuitively, this is particularly useful for derivatives that rely on
rare events. A deep out-of-the money option is an obvious example for rare
event dependency. Crude MC would only rarely produce samples which lead to
non-zero payouts and, consequently, the MC variance would be severe. However,
IS is by far not limited to rare event cases. 
Compared to other VR techniques the usage of IS is 
more involved, because the selection of a suitable proposal is generally difficult.
But the additional effort is justified by the large potential of IS to reduce
the MC variance.

IS has been successfully applied to derivative pricing
 based on Gaussian proposals. That is, the proposal was chosen from
some class of Gaussian distributions. An important approach is based on a mean
shift, which can be obtained through saddle point approximation (Glasserman,
Heidelberger, and Shahabuddin 1999), adaptive stochastic optimization
(Vazquez-Abad and Dufresne 1998; Su and Fu 2000, 2002), or least squares
(Capriotti 2008). This approach is also known as the ``change-of-drift technique''.
In addition, Gaussian mixture distributions have been utilized for
approximating the optimal proposal (Avramidis 2002).
Summarizing, existing approaches are based on parametric IS,
that is the proposal is chosen from a certain class of distributions. 
For complex payouts it is hard to set up a class which contains a distribution
that approximates the optimal proposal reasonably well.

This paper proposes the usage of nonparametric IS (NIS) for
derivative pricing. 
The basic idea of NIS is to use a nonparametric estimate of the optimal
proposal. NIS algorithms have already been successfully applied to
low-dimensional integration problems (Givens and Raftery 1996; Zhang 1996;
Kim, Roh, and Lee 2000; Neddermeyer 2009). However,
high-dimensional integration tasks have not been considered until now.
As a result of the curse of dimensionality and computational limitations 
NIS cannot be applied directly to high-dimensional derivative pricing.
We suggest to restrict NIS to those coordinates that are of most
importance to the integration problem. This approach can be justified by the concept of the
effective dimension (ED). To reduce the effective dimension and to identify the
most relevant coordinates principal component analysis (PCA) is applied.

The advantage of NIS compared to parametric IS is its close approximation of
the optimal proposal. We prove that the VR factor of our
nonparametric method increases with sample size converging to the \-- in some
sense \-- optimal value. 
Note, parametric IS methods achieve constant VR factors.
It is shown through simulations that the proposed algorithm is computationally
more efficient than parametric IS for well-known benchmark option
pricing problems. In the case of low ED, the algorithm does not only
outperform in terms of MSE but also in terms of computational costs. In other
words, it is not only more accurate but also computationally cheaper. In
addition, it is easy to implement based on the C$++$ implementation of the
nonparametric estimator which is provided.
NIS and most parametric IS methods share the property that they can be 
combined with other VR techniques. This is demonstrated through 
the use of low-discrepancy sequences (also known as quasi-MC (QMC)).

This paper is organized as follows.
In the next section, a general MC option pricing
framework is introduced and IS is briefly reviewed. In Section 3, a
nonparametric partial IS algorithm is proposed and its MSE
convergence properties are investigated. 
The concept of the ED is reviewed in
Section 4. In Section 5, path construction based on PCA is
discussed, followed by a brief introduction to QMC (Section 6). 
A comparison to parametric IS is presented in Section 7 and a detailed
description of the implementation is given in Section 8. Finally, in Section
9 simulation results are reported followed by conclusions (Section 10).

\newpage

\vspace{0.5cm}

\begin{center}
2. DERIVATIVE PRICING AND IMPORTANCE SAMPLING
\end{center}

Let's describe the evolution of the underlying asset through a stochastic
differential equation (SDE) of the form
\begin{equation}\label{op:sde}
d S(t) = r S(t)\ dt + \sigma(S(t)) S(t) dW(t),
\end{equation}
where $W(t)$ is a standard Brownian motion (BM);  
$r$ and $\sigma$ are the risk-free interest rate and the volatility,
respectively. Within this model, evaluating the price of a European option with
payout function $C_K(S)$, strike level $K$, and expiry $T$ means computing
\begin{equation}\label{op:exp}
\mathbf{E}[\exp(-rT) C_K(S)],
\end{equation}
where the expectation is taken with respect to the risk neutral measure. Except
of special cases, there is no explicit solution for SDEs of kind
(\ref{op:sde}). Therefore, it is required to migrate to some discretization
$\tilde{S}_{t_k}$ of the process $S(t)$, which is defined on a discrete time
grid $0=t_0<t_1 < \dots < t_{d} = T$. The first order Euler discretization scheme yields
\begin{equation}\label{eulerpath}
\tilde{S}_{t_{k+1}} = \tilde{S}_{t_k} + r \tilde{S}_{t_k} (t_{k+1}-t_k) +
\sigma(\tilde{S}_{t_k}) \tilde{S}_{t_k} \sqrt{t_{k+1}-t_k} Z_{t_k}
\end{equation}
with standard normal innovations $Z_{t_k}$. 
In the following, we focus on an equally-spaced time grid, i.\ e. $t_i-t_{i-1}
= \Delta t = \text{const}$. Based on this discretization, the option price
(\ref{op:exp}) can be approximated through the integral
\begin{equation*}
I_{\varphi} = \int_{\RR^d} \varphi(\mathbf{x}) p(\mathbf{x}) d \mathbf{x},
\end{equation*}
where $\varphi(\mathbf{x}) = \exp(-rT)
C_K(\tilde{S}(\mathbf{x}))$. $p$ denotes the density of the
multivariate Gaussian distribution $\mathcal{N}(\mathbf{0},\mathbf{I}_d)$ with $\mathbf{I}_d$
being the identity matrix of dimension $d$. By writing $\tilde{S}(\mathbf{x})$,
it is meant that a trajectory of $\tilde{S}_{t_k}$ is built up based on
the innovations $\mathbf{x} = (x_1, \ldots, x_d)^T$. To keep the discretization
bias small, it is required to choose $d$ considerably large which leads to a high-dimensional integration problem.
The crude MC estimator is given by
\begin{equation*}
\hat{I}_{\varphi}^{\text{MC}} = \frac1N \sum_{i=1}^N \varphi(\mathbf{x}^i),
\end{equation*}
where $\mathbf{x}^1, \mathbf{x}^2, \ldots, \mathbf{x}^N$ are drawn from $p$. 
The estimator's large variance renders it impractical for
approximating complex integrals.
To construct a more efficient MC estimator, IS can be applied.
The basic idea of IS is to sample from a proposal $q$
instead of $p$.  
The IS estimator is defined through
\begin{equation*}
\hat{I}_{\varphi}^{\text{IS}} = \frac1N \sum_{i=1}^{N} \varphi(\mathbf{x}^i)
w(\mathbf{x}^i),
\end{equation*}
with likelihood ratio $w(\mathbf{x}) = p(\mathbf{x}) / q(\mathbf{x})$ and
samples $\mathbf{x}^1, \mathbf{x}^2, \ldots, \mathbf{x}^N$ drawn from $q$. 
For $\hat{I}_{\varphi}^{\text{IS}}$ to converge to $I_{\varphi}$, it is
required that the support of $q$ includes the support of $|\varphi|p$. Under
the additional assumption $\text{Var}_q[\varphi w] < \infty$, a central limit
theorem holds
\begin{equation*}
\sqrt{N} (\hat{I}_{\varphi}^{\text{IS}} - I_{\varphi}) \Rightarrow
\mathcal{N}(0, \sigma^2_{\text{IS}}),
\end{equation*}
where $\sigma^2_{\text{IS}} = \mathbf{E}_q[\varphi
w-I_{\varphi}]^2$ (Rubinstein 1981). 
This result allows the construction of confidence intervals for
$\hat{I}_{\varphi}^{\text{IS}}$ and establishes the convergence rate
$\mathcal{O}(N^{-1/2})$. 
The optimal proposal, which minimizes the (asymptotic)
variance $\sigma^2_{\text{IS}}$, is given by
\begin{equation*}\label{is:optdis}
q^{\text{opt}}(\mathbf{x}) = \frac{|\varphi(\mathbf{x})|p(\mathbf{x})}
{\int |\varphi(\mathbf{x})| p(\mathbf{x}) d\mathbf{x}}.
\end{equation*}
Remarkably, the IS estimator based on the optimal proposal $q^{\text{opt}}$
has zero variance for functions $\varphi$ with a definite sign. 
However, the optimal proposal is unavailable in practice because of its 
unknown denominator.

\vspace{0.5cm}

\begin{center}
3. NONPARAMETRIC PARTIAL IMPORTANCE SAMPLING
\end{center}

In this section, nonparametric partial IS (NPIS) is introduced as a
generalization of the NIS algorithm discussed in Zhang (1996)
and Neddermeyer (2009). NIS is a two-stage procedure. In the first
stage, the optimal proposal is estimated nonparametrically based on samples
drawn from a trial distribution $q_0$. In the second stage,  this nonparametric
density estimate is used as proposal for IS. 
We pick up this approach, but instead of approximating the optimal proposal in
the entire space, we focus on the optimal proposal in a certain subspace. That
is, the nonparametric IS procedure is restricted to a low-dimensional
subproblem in order to avoid the curse of dimensionality. 
We decompose $\mathbf{x} = (\mathbf{x}_u, \mathbf{x}_{-u})$, where {${u
\subseteq \{1,2,\ldots,d\}}$}. The cardinality of $u$ is denoted by $|u|$. 
Let's consider the marginalized optimal proposal obtained through integration
with respect to $\mathbf{x}_{-u}$. It is given by
\begin{equation*}
\breve{q}^{\text{opt}} (\mathbf{x}_{u}) = \int_{\RR^{d-|u|}}
q^{\text{opt}}(\mathbf{x}) d \mathbf{x}_{-u}.
\end{equation*}
Subspace $u$ is chosen such that it covers those coordinates which are most
important to the integrand (see Section 4). 
To limit the computational burden of the nonparametric method, $u$ will be
considerably small in practice ($1 \leq |u| \leq 3$).
In the NPIS algorithm, $\breve{q}^{\text{opt}}$ is estimated
nonparametrically. Now, the choice of the nonparametric estimator is discussed.
 Typically, kernel estimators have been used within NIS algorithms. However,
 sampling and evaluation is computationally very inefficient for kernel
 estimators. In this article, a multivariate variant of the nonparametric
 frequency polygon estimator is used, which is known as linear blend frequency
 polygon (LBFP) (Terrell 1983). The LBFP estimator attains the same MSE
 convergence rate as kernel estimators, namely $\mathcal{O}(N^{-4/(4+d)})$, 
 but it is computationally more efficient. 
 The generation and the evaluation of $N$ samples is of order
 $\mathcal{O}(2^{d-1} d^2 N^{(d+5)/(d+4)})$ (Neddermeyer 2009), whereas kernel estimators require
 $\mathcal{O}(dN^2)$ operations. The construction of an LBFP consists of
 interpolations of the bin mid-points of a multivariate histogram. 
 Let's consider a multivariate histogram estimator with bin height
$\hat{f}^{\text{H}}_{k_1, \ldots, k_d}$ for bin $B_{k_1, \ldots, k_d} =
\prod_{i=1}^d [t_{k_i}-h/2, t_{k_i} + h/2)$, where $h$ is the bin width and
$(t_{k_1}, \ldots, t_{k_d})$ the bin mid-point.
For ${\bf x} \in \prod_{i=1}^{d} [t_{k_i},t_{k_i} + h)$ the LBFP estimator is
defined as
\begin{equation}\label{def:LBFP}
\hat{f} ({\bf x}) = \sum_{j_1, \ldots, j_d \in \{0,1\}} \left[
\prod_{i=1}^{d} \left(\frac{x_i-t_{k_i}}{h}\right)^{j_i}
\left(1-\frac{x_i-t_{k_i}}{h}\right)^{1-j_i} \right]
\hat{f}^{\text{H}}_{k_1+j_1,\ldots,k_d+j_d}.
\end{equation}
In the one-dimensional case, $\hat{f}$ is just a linearly interpolated
histogram. It can be shown that it integrates to one.

\bigskip

\hrule

\medskip

\noindent
\textbf{Algorithm \-- Nonparametric Partial Importance Sampling}

\vskip 2mm

\noindent
{\em Stage 1: Nonparametric estimation of the marginalized optimal proposal}
	\begin{itemize}
		\item Select subset $u$, bin width $h$, trial distribution $q_0$, and sample
		sizes $M$ and $N$.
		\item {\bf For $j=1,\ldots,M$}: $\;$ Generate sample $\tilde{\mathbf{x}}^j
		\sim q_0$.
		\item Obtain nonparametric estimate $\hat{q}^{\text{opt}}$ of
		marginalized optimal proposal $\breve{q}^{\text{opt}}$
		\[ \hat{q}^{\text{opt}}(\mathbf{x}_{u}) = 
        \frac{\hat{f}(\mathbf{x}_{u})} {\frac1M \sum_{j=1}^M \omega^j},
		\] 
		where $\omega^j =
		|\varphi(\tilde{\mathbf{x}}^j)| p(\tilde{\mathbf{x}}^j)
		q_0(\tilde{\mathbf{x}}^j)^{-1}$ and
		\begin{eqnarray*}
        \hat{f}(\mathbf{x}_{u}) &=& \frac1M  \sum_{j_1, \ldots, j_{|u|} \in
        \{0,1\}} \left[
\prod_{i \in u} \left(\frac{x_i - t_{k_i}}{h}\right)^{j_i}
\left(1-\frac{x_i - t_{k_i}}{h}\right)^{1-j_i} \right] \\
		&& \qquad \times  \sum_{j=1}^{M} \omega^j \mathbf{1}_{\prod_{i \in u}
[t_{k_i+j_i} - h/2,t_{k_i+j_i} + h/2)}(\tilde{\mathbf{x}}^j)
		\end{eqnarray*}
		for $\mathbf{x}_{u} \in \prod_{i \in u} [t_{k_i},t_{k_i} + h)$.
		\end{itemize}
	{\em Stage 2: Partial Importance Sampling}
	\begin{itemize}
		\item {\bf For $i=1,\ldots,N$}: $\;$ Generate samples $\mathbf{x}_{u}^i \sim
		\hat{q}^{\text{opt}}(\mathbf{x}_{u})$ and $\mathbf{x}_{-u}^i \sim p(\mathbf{x}_{-u})$.
		\item Evaluate 
		\[ \hat{I}_{\varphi}^{\text{NPIS}} =
		\frac{1}{N} \sum_{i=1}^{N} \varphi(\mathbf{x}^i) p(\mathbf{x}_{u}^i)
		\hat{q}^{\text{opt}}(\mathbf{x}_{u}^i)^{-1}.\]
	\end{itemize}
\hrule

\medskip
\noindent
The following theorem investigates the MSE convergence properties of the NPIS
algorithm to obtain the optimal value for bin width $h$.
\begin{thm}\label{thm:npis}
Suppose that the assumptions given in Appendix A hold, $\varphi \geq
0$, and $p(\mathbf{x}) = p(\mathbf{x}_{u}) p(\mathbf{x}_{-u})$. We denote $\breve{q} =
\breve{q}^{\text{opt}}$.
Then, we obtain for $\hat{I}_{\varphi_M}^{\text{NPIS}}$ (as defined in Appendix
A)
\begin{equation}\label{npis:mse}
\mathbf{E}[\hat{I}_{\varphi_M}^{\text{NPIS}} -
I_{\varphi}]^2 =
\frac{1}{N} \left[ \int \frac{\nu(\mathbf{x})^2 p(\mathbf{x}_{-u})}
{\breve{q}(\mathbf{x}_{u})} d\mathbf{x} \ + \ I_{\varphi}^2 \left\{ h^4 H_1 +
\frac{2^{|u|}}{3^{|u|} M h^{|u|}} H_2 \right\} \times (1 + o(1)) \right]
\end{equation}
and the optimal bin width
\[ h^{\text{opt}} = \left(\frac{{|u|} H_2 2^{|u|}}{4 H_1
3^{|u|} }\right)^{\frac{1}{4+|u|}} M^{-\frac{1}{4+|u|}}, \]
where 
\[H_1 = \frac{49}{2,880} \sum_{i \in u} \int
\frac{(\partial_i^2\breve{q})^2}{\breve{q}} + \frac{1}{64} \sum_{i,j \in
u\atop {i\not= j}} \int \frac{\partial_i^2\breve{q}
\partial_j^2\breve{q}}{\breve{q}}, \quad H_2 = \int \frac{(q^{\text{opt}})^2}{\breve{q} \ q_0 }\]
and 
\[\nu(\mathbf{x}) = \varphi(\mathbf{x}) p(\mathbf{x}_{u}) - \int
\varphi(\mathbf{x}) p(\mathbf{x}) d\mathbf{x}_{-u}.\]
\end{thm}
\proof See Appendix A.

The left and right term in the brackets in (\ref{npis:mse}) can be interpreted
as the variance caused by the components $\mathbf{x}_{-u}$ and $\mathbf{x}_{u}$,
respectively. Note, subset $u$ is chosen such that the left term is small
compared to the right one. The expression in braces quantifies the MSE of the
nonparametric estimate, which depends on both $\breve{q}^{\text{opt}}$ and trial
distribution $q_0$. For $h = h^{\text{opt}}$ and $M/N \rightarrow \lambda \in
(0,1) \; \; (M,N \rightarrow \infty)$
the theorem implies
\begin{equation*}
\mathbf{E}[\hat{I}_{\varphi_M}^{\text{NPIS}} - I_{\varphi}]^2 = \frac{1}{N} \int
\frac{\nu(\mathbf{x})^2 p(\mathbf{x}_{-u})} {\breve{q}(\mathbf{x}_{u})}
d\mathbf{x} + \mathcal{O}(N^{-(8+|u|)/(4+|u|)}).
\end{equation*}
Hence, the variance caused by $\mathbf{x}_u$ is of lower order. In other
words, the optimal variance (for partial IS on coordinates $u$) is achieved
asymptotically.
As a consequence, compared to crude MC and parametric IS
techniques, NPIS is expected to yield increasing efficiency as the sample size
grows. Furthermore, if $|u|=d$ the MSE converges as fast as
$\mathcal{O}(N^{-(8+d)/(4+d)})$, which is a massive improvement compared to
the standard MC rate $\mathcal{O}(N^{-1})$, for $d$ that is small (Zhang 1996;
Neddermeyer 2009).
Note, the results of this section also hold for distributions $p$ other than the
standard normal distribution.

In this article, NPIS is only investigated for non-negative integrands.
However, by decomposing the payout function $C = C^{+} - C^{-}$, NPIS can also
be applied to financial derivatives that have both positive and negative
payouts.

\vspace{0.5cm}

\begin{center}
4. EFFECTIVE DIMENSION
\end{center}

The NPIS algorithm is based on the restriction on specific
coordinates $\mathbf{x}_u$, where in high-dimensional integration
problems $|u|\ll d$. This approach can be justified by the concept of
the ED. It is well known, that many integration problems in
financial engineering, despite having a large nominal dimension, are 
low-dimensional in terms of the ED. 
For a rigorous definition of the ED, let's consider the
functional analysis of variance (ANOVA) decomposition. 
Suppose $\int \varphi(\mathbf{x})^2 p(\mathbf{x})d\mathbf{x} < \infty$ and
$p(\mathbf{x}) = \prod_{i=1}^{d} p(\mathbf{x}_i)$ is a product
density. Then, $\varphi$ can be written as a sum of $2^d$ orthogonal
functions
\[
\varphi(\mathbf{x}) = \sum_{u \subseteq \{1,2,\ldots,d\}}
\varphi_u(\mathbf{x}_u),
\]
where the ANOVA functions $\varphi_u$ are given recursively by
\[
\varphi_u(\mathbf{x}_u) = \int_{\RR^{d-|u|}} \varphi(\mathbf{x})
p(\mathbf{x}_{-u}) d \mathbf{x}_{-u} - \sum_{v \subset u}
\varphi_v(\mathbf{x}_v).
\] 
Now, the fraction of the variance $\sigma^2 =
\text{Var}_p[\varphi]$, that is explained by certain lower-dimensional ANOVA
functions, is considered. For this purpose, the variance of
$\varphi_u$ is defined by $\sigma^2_u = \int_{\RR^d}
\varphi_u(\mathbf{x}_u)^2 p(\mathbf{x}) d\mathbf{x}$, where 
$\sigma^2_{\emptyset} = 0$. As the ANOVA decomposition is orthogonal, one has
$\sigma^2 = \sum_u \sigma^2_u$. Hence, $\Gamma_u =\sum_{v \subseteq u}
\sigma^2_v$ can be interpreted as the contribution of $\mathbf{x}_u$ to the
total variance of $\varphi$. For a more detailed description of the ANOVA decomposition
see, for instance, Takemura (1983) and Owen (1992). The following
definition of the ED is due to Caflisch, Morokoff, and Owen (1997).
\begin{defn}
The ED (in the truncation sense) is the cardinality of the
smallest subset $u$ such that $\Gamma_u \geq \gamma \sigma^2$
with $0<\gamma <1$.
\end{defn}
The threshold $\gamma$ is chosen close to one. In the framework of this article,
we found $\gamma = 0.9$ reasonable.
The ED does not only allow to identify those coordinates
which most effect the integral value but it also indicates how many coordinates
are required to cover a certain amount of the variance.
An MC procedure that allows one to determine the ED of a given problem is
described in Appendix B.

\vspace{0.5cm}

\begin{center}
5. GAUSSIAN MODELS
\end{center}

The purpose of this section is to show how NPIS can be applied to models that
are based on the integration with respect to high-dimensional Gaussian
distributions. 
As mentioned before, NPIS is inefficient as a result of the curse of
dimensionality unless the ED (and thus $|u|$) is small. 
For typical financial integration problems it is generally not advisable to
apply NPIS with $|u|$ larger than 3, unless the number of paths to be sampled is
huge or the domain of interest is very small (rare event case). 

Suppose the task is to integrate with respect to $\mathcal{N}(0,\Sigma)$. 
Now, PCA is used to transform the problem.
The (positive-definite) covariance matrix $\Sigma$ is written as
\begin{equation*}
\Sigma = V \Lambda V^T,
\end{equation*}
with $\Lambda = \text{diag}(\lambda_1, \lambda_2, \ldots, \lambda_d)$ and 
eigenvalues $\lambda_i$. The columns of $V$ are the corresponding unit-length
eigenvectors. Thus, one has $V \Lambda^{1/2} \mathbf{Z} \sim
\mathcal{N}(0,\Sigma)$ for $\mathbf{Z} \sim \mathcal{N}(0,I_d)$.
Without loss of generality, it is assumed that the eigenvalues (and
the corresponding eigenvectors) are sorted so that $\lambda_1 \geq \lambda_2
\geq \cdots \geq \lambda_d$. 
The PCA construction of samples from $\mathcal{N}(0,\Sigma)$ is optimal in the
sense that it provides an optimal lower-dimensional approximation (in the MSE sense) to the
random variable of interest. 
This means that the first $k$ components of $\mathbf{Z}$ explain as much as
possible of the total variance. More precisely, it can be shown that they
explain the fraction $ (\lambda_1+ \lambda_2+ \ldots+ \lambda_k)/(\lambda_{1}+
\lambda_{2}+ \ldots + \lambda_d)$.

\begin{figure}
\centering
\includegraphics[width=450pt, keepaspectratio]{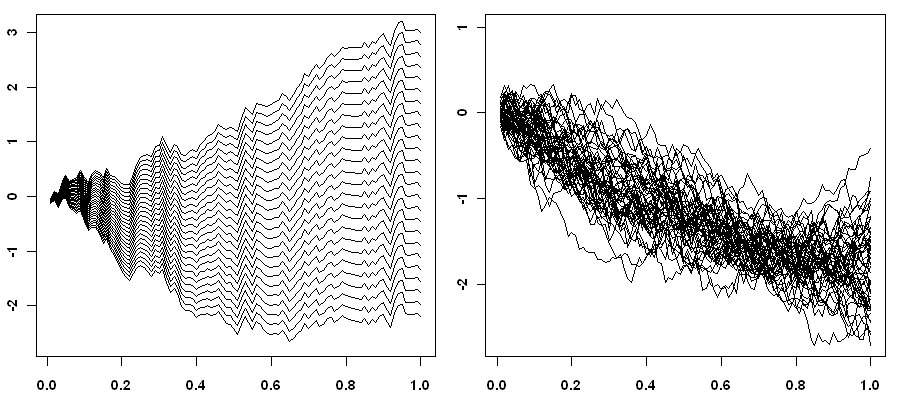}
\caption{\footnotesize Discretized BM paths: first PCA
component varies whereas other components are fixed to random values (left);
first PCA component is fixed and other components vary randomly (right).}
\label{fig:PCAplots_neu}
\end{figure}

The option pricing problem introduced in Section 2 leads
to the construction of discretized BM paths based on samples from
the multivariate Gaussian distribution. Paths are most easily build up through the
random walk construction guided by (\ref{eulerpath}). In this construction each
component ``counts roughly the same'' rendering the restriction on a
lower-dimensional subspace and hence the application of NPIS impractical.
Note, the integral $I_{\varphi}$ can be rewritten as $I_{\varphi} = \int
\tilde{\varphi}(\mathbf{x}) p_{\mathcal{N}(\mathbf{0}, \Sigma)}(\mathbf{x})
d\mathbf{x}$, where $\Sigma$ is the covariance matrix of the discretized BM
with entries $\Sigma_{ij} = \min\{ t_i, t_j \}$. 
This suggests that PCA can be used to reduce the ED.
The PCA construction of discretized BM paths has a continuous limit known as
Karhounen-Lo{\`{e}}ve expansion of BM:
\begin{equation*}
W(t) = \sum_{i=1}^{\infty} \sqrt{\lambda_i} \psi_i(t) Z_i, \qquad 0 \leq
t \leq 1,
\end{equation*}
where $\psi_i(t) = \sqrt{2} \sin \{(i-0.5) \pi t\}$,
$\lambda_i = \{(i-0.5)\pi \}^{-2}$, and $Z_i \sim
\mathcal{N}(0,1)$ (Adler 1990).
Based on the expression for $\lambda_i$, it is easily shown that $Z_i$ 
explains the fraction $2 \lambda_i$ of the path's variability (which is
approximately 81\%, 9\%, 3\% for $i=1,2,3$, respectively). These values are not
only of asymptotic nature but also hold for a small number of discretization steps (with slight
deviations). This astonishing result claims that very few PCA
components suffice to determine most of the path's variation no matter how long or
detailed it is. 
Particularly, the first PCA component plays a dominant role and has a nice
geometrical interpretation. Roughly speaking, it determines the path's direction in the path space. 
This is visualized in Figure \ref{fig:PCAplots_neu}. 

Another common method for the reduction of the ED (of a discretized BM) is
the Brownian Bridge technique. In this paper, the focus is on
PCA because of its optimality property. However, it is remarked that in
certain situations Brownian Bridge techniques are superior to PCA. This may
especially be the case if the payout function only depends on the terminal
value of the underlying. Note, NPIS can also be combined with Brownian Bridge techniques.

\vspace{0.5cm}

\begin{center}
6. QUASI-MONTE CARLO INTEGRATION
\end{center}

QMC is often used to (further) improve MC estimators.
In contrast to MC, QMC integration uses
so-called low-discrepancy sequences instead of (pseudo-) random numbers.
Low-discrepancy numbers are constructed to fill the
space more evenly.
For a description of the construction of low-discrepancy sequences readers are
referred to Niederreiter (1992), Glasserman (2004), and the references given
there.
The incentive to work with QMC is justified by its
deterministic error bound of order $\mathcal{O}(N^{-1} \log^d N)$, which follows
from the well-known Koksma-Hlawka inequality (see Niederreiter (1992)).
This bound is merely of theoretical benefit because the
computation of the involved constants (including the Hardy-Krause
variation of the integrand) is infeasible or at least very
difficult.
However, it suggests that QMC should
massively outperform MC in low-dimensional integration problems. 
The advantage of QMC diminishes with increasing dimension.
Nevertheless, it is well known in the financial engineering literature, that QMC
may be effectively applied to high-dimensional problems (Paskov and Traub
1995; Ninomiya and Tezuka 1996; Traub and Werschulz 1998). 
This stems from the fact mentioned earlier that many problems in finance have
rather low ED compared to the nominal dimension.
As the convergence properties of QMC become worse in higher
dimensions, it is important to assign the first coordinates to the most relevant
dimensions of the integration problem. In our setting, the relevant
coordinates are those contained in $u$.

A drawback of QMC is the lack of randomness, which impedes the
computation of the MSE for assessing the accuracy of the estimator. 
This issue can be resolved by randomizing the deterministic low-discrepancy
sequence to achieve independent realizations of the QMC estimator. 
Different approaches for randomizing low-discrepancy sequences are available
including Owen's scrambling (Owen 1995), random digit scrambling
(Matou{\v{s}}ek 1998), or random shifts (see {\"O}kten and Eastman (2004) for a survey). 
In our simulations, priority is given to the random shift technique
because of its straightforward implementation. The idea is to shift the entire sequence by a random vector $\mathbf{v}$
modulo one. $\mathbf{v}$ is drawn from the uniform distribution on $[0,1)^d$. 
That is, a randomized sequence is obtained by substituting the quasi-random 
vectors $\mathbf{y}^i$  of the original low-discrepancy sequence by
$(\mathbf{y}^i + \mathbf{v}) \ \text{mod} \ 1$.

\vspace{0.5cm}

\begin{center}
7. COMPARISON TO PARAMETRIC IMPORTANCE SAMPLING
\end{center}

Until now, the application of IS in finance was limited to parametric
IS. In particular, Gaussian or mixtures of Gaussian distributions have been
applied.
The variance of a parametric IS estimator with proposal $q_{\theta}$ (and
parameter $\theta \in \Theta$) can be written as
\begin{equation}\label{var:paramIS}
\frac{\sigma^2_{\text{IS}}} {N} = \frac{I^2_{\varphi}}{N} \left\{ \int_{\RR^d}
\frac{q^{\text{opt}}(\mathbf{x})^2} {q_{\theta}(\mathbf{x})} d\mathbf{x} -1
\right\},
\end{equation}
where $\sigma^2_{\text{IS}}$ is defined as in Section 2.
First, this suggests that, in contrast to NPIS, the VR factor is
constant because all terms are $\mathcal{O}(N^{-1})$. Second, the variance is critically affected by
the tails of $q_{\theta}$. Using
Gaussian proposals, it is often hard to approximate the tails of
$q^{\text{opt}}$ reasonably well. There lies a distinct advantage of NIS
methods. Most parametric IS approaches aim at choosing $\theta$ so that
(\ref{var:paramIS}) is minimized. 
We now discuss a variant of the least-squares IS (LSIS) algorithm (Capriotti
2008) which is directly comparable to NPIS. It is based on the Gaussian
proposal $\mathcal{N}(\mu, \mathbf{I}_d)$ with parameter $\mu \in \RR^d$.
Similar to NPIS, it is a two-stage algorithm. In the first stage, based on $M$
samples from $p$, a least-squares problem is solved to estimate the optimal
drift change $\mu$. (The variance can also be adjusted through this procedure.) 
However, as the problem dimension grows the estimate of $\mu$ becomes unreliable. 
The variant of this algorithm which is suggested here applies LSIS to the
coordinates $\mathbf{x}_u$, that are determined
through PCA and the ED (analogous to NPIS). This makes the LSIS and the NPIS
directly comparable. In Section 9, NPIS and this variant of LSIS are tested
against each other through simulations.

Besides the superior convergence properties NPIS has a computational advantage
over parametric IS which is of relevance in practice. For
computing the IS weights, parametric IS typically needs to evaluate the \texttt{exp} function
which is very expensive. Through the use of the LBFP estimator, these
evaluations are reduced in the NPIS algorithm. This leads to a relevant
reduction of the computational costs (compare Section 9).
 
Finally, we remark that combinations of parametric IS and NPIS are possible.
For instance, while applying NPIS to $\mathbf{x}_{u}$ one can carry out
parametric IS on the remaining coordinates $\mathbf{x}_{-u}$.

\vspace{0.5cm}

\begin{center}
8. IMPLEMENTATION OF THE ALGORITHM
\end{center}

In this section, the details of practical implementation of the proposed NPIS
algorithm is discussed. 
At first, an overview over the
required ingredients for the implementation is given.

\vspace{0.5cm}

\begin{center}
8.1 OVERVIEW
\end{center}

\begin{description}
  \item[Subset $u$:] This is chosen according to the ED ($\gamma=0.9$), 
  which can be computed with the algorithm given in Appendix B. If PCA is
  used, the first few principal components are selected.
  \item[Trial distribution $q_0$:] The choice of the trial distribution should
  be guided by the following two criteria: First, it should allow for  efficient
  sampling and evaluation. Second, the marginal distributions of the
  coordinates contained in $u$ should be overdispersed (heavy tailed) compared
  to the standard normal distribution. An all-purpose trial distribution,
  which we found to work well in practice, is discussed in Subsection 8.2. 
  Alternatively, one can use a parametric choice tailored to the specific
  integration problem or one can simply use the (multivariate) standard normal
  distribution. The latter is often not a good choice because of
  the importance of the tails of the proposal.
  \item[Bin width $h$:] A Gaussian reference rule can be computed in
  Stage 1 of the algorithm (for details see Subsection 8.3).
  \item[Sample size $M$:] For the simulations, we used $M = \max\{256,
  0.25N\}$. In the special case when $|u|=d$ an optimal value for the
  proportion $M/N$ can be derived (Zhang 1996; Neddermeyer 2009).
  \item[LBFP estimator:] The details of the implementation of the LBFP
  estimator can be found in Neddermeyer (2009). A C$++$ implementation of the
  LBFP as well as the R-package lbfp are available on request.
\end{description}

We emphasize that, in contrast to most parametric IS algorithms, all parameters
are adjusted automatically, such that no trial-and-error parameter selection
and no analytical computation are necessary in practice.

\vspace{0.5cm}

\begin{center}
8.2 TRIAL DISTRIBUTION
\end{center}

As trial distribution we propose a simple product density.
It is composed of a uniform distribution on $[-\rho_M, \rho_M]^{|u|}$ and the
multivariate Gaussian distribution $p(\mathbf{x}_{-u})$:
\begin{equation*}
q_0(\mathbf{x}) = p(\mathbf{x}_{-u}) \times \frac{1}{(2 \rho_M)^{|u|}}  \prod_{i
\in u} \mathbf{1}_{[-\rho_{M}, \rho_{M}]}(x_i),
\end{equation*}
where $\rho_{M}$ is the $(1+ (1-\epsilon)^{1/M}) / 2$-quantile of
$\mathcal{N}(0,1)$. $\epsilon>0$ is very small, say $\epsilon =
10^{-4}$. Consequently, $\mathbf{P}(\max_{1\leq i \leq M} |Z_i| >
\rho_{M}) = \epsilon$ holds for standard normal distributed $Z_i$. This ensures
that the bias caused by the bounded support of the uniform distribution is very
small. In addition, the uniform distribution guaranties that the space of
$\mathbf{x}_{u}$ is well explored even for a small sample size.

\vspace{0.5cm}

\begin{center}
8.3 PRACTICAL BIN WIDTH SELECTION
\end{center}

The expression for $h^{\text{opt}}$ given in Theorem
\ref{thm:npis} is intractable analytically because of the unknown constants
$H_1$, $H_2$. The plug-in method suggested in Zhang (1996) also seems
unsuitable for our integration problem as derivatives of the integrand are
required. We propose to apply a Gaussian approximation of $H_1$, $H_2$.
Suppose $\breve{q}^{\text{opt}}$ is the density of a centered multivariate
Gaussian distribution with covariance matrix 
$\text{diag}(\sigma_1^2,\sigma_2^2, \ldots, \sigma_{|u|}^2)$. Under this
assumption, it can be shown that
\begin{equation}\label{simpleH1}
H_1 = \frac{98}{2,880} \sum_{i \in u} \sigma_i^{-4}.
\end{equation}
For the constant $H_2$ the mean of $q^{\text{opt}}$ plays the dominant role.
Therefore, it is assumed that $q^{\text{opt}}$ is the density of
$\mathcal{N}((\mu_1, \mu_2, \ldots, \mu_{d})^T, \mathbf{I}_{d})$. If the trial
distribution is chosen as explained in the preceding subsection, one yields
\begin{equation}\label{simpleH2}
H_2 \approx \rho_M^{|u|} \exp[\sum_{i \notin u} \mu_i^2].
\end{equation}
In the algorithm, the expressions in (\ref{simpleH1}) and (\ref{simpleH2}) can
be approximated based on the samples
$\tilde{\mathbf{x}}^1, \ldots, \tilde{\mathbf{x}}^M$. 
This follows from the fact, that the samples $\tilde{\mathbf{x}}^j$ weighted
with $\omega^j/\sum_{k=1}^{M} \omega^k$ approximate $q^{\text{opt}}$.

\vspace{0.5cm}

\begin{center}
9. SIMULATION RESULTS
\end{center}

Different European option pricing scenarios are considered to compare the
proposed algorithms (NPIS and the combination of NPIS and QMC (QNPIS))
with existing methods (crude MC, QMC, LSIS, and the combination of LSIS and QMC
(QLSIS)).
The performance of the algorithms
is measured through the VR factors (computed with respect
to crude MC) and the relative computational efficiency (RCE). 
The RCE is defined as the ratio of the CE
of the method of interest to the CE of crude MC.
The computational costs are measured in seconds. 
All simulations are done for different sample sizes $N$ in order to demonstrate
the increasing VR factors of NPIS.

Examples 1 through 3 consider different single- and multi-asset options 
within the standard Black-Scholes model. There, the price of an asset
$S$ at time $t$ is given by 
\begin{equation*} 
S(t) = S(0) \exp[(r-0.5 \sigma^2) t + \sigma \sqrt{t} Z]
\end{equation*}
with standard normal random variable $Z$. 
The simulations are based on the following setting:
$S(0) =$ 100, $\sigma = 0.3$, $r=0.05$, and $T=1$. 
In Example 4, the pricing of a cap within the CIR
model is investigated to show the effectiveness of NPIS/QNPIS in a
square-root diffusion model.
For all algorithms, apart from crude MC, the PCA path
construction is used. The parameters $u$, $q_0$, and $h$ are chosen according to
the description in the preceding section.
Note, Theorem 1 does not apply to QMC sampling. We found empirically that QNPIS
requires a larger bin width. In the simulations, $3h^{\text{opt}}$ is used.
For LSIS and NPIS, $M$ is set as suggested in the preceding section whereas for
QNPIS and QLSIS $M=$ 1024 is used throughout.
The least squares estimates required in LSIS/QLSIS were computed with ten
iterations of the Levenberg-Marquardt method (Press et al. 1992, pp. 683-688).

The computations were carried out on a Dell Precision T3400,
Intel CPU 2.83GHz. All algorithms were coded in C$++$. The
Mersenne Twister 19937 (Matsumoto and Nishimura 1998) and the Sobol sequence
(Sobol 1967) were used for pseudo- and quasi-random number generation,
respectively. The Sobol sequence is randomized by the random shift
technique. The transformation of uniform random numbers into normal random
numbers was done by the Beasley-Springer-Moro approximation (Moro 1995).

\vspace{0.5cm}

\begin{center}
Example 1. Straddle Option
\end{center}

\begin{figure}
\centering
\includegraphics[height=200pt, keepaspectratio]{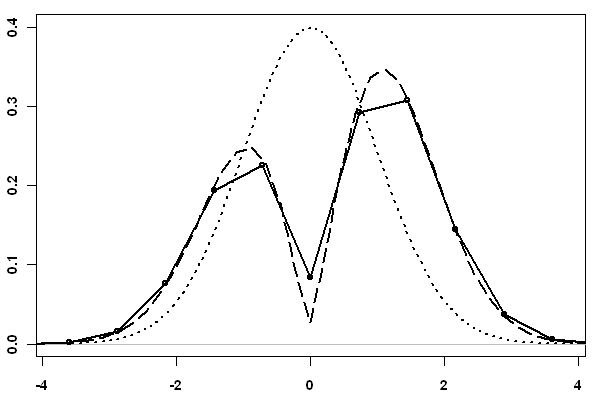}
\caption{\footnotesize Standard normal distribution (dotted line),
optimal proposal for a straddle option within the Black-Scholes
model (dashed line), and an LBFP estimate of the optimal proposal (solid line).
Model parameters: $S(0) =$ 100, $\sigma = 0.3$, $r=0.05$, $T=1$, and $K=$ 100.}
\label{fig:StraddleOptProp}
\end{figure}

The payout function of a straddle option is given by
\[
C_K(S) = (S(T)-K)^{+} + (K-S(T))^{+}.
\]
In the Black-Scholes world the pricing of a straddle option is a 
one-dimensional integration problem with multi-modal optimal proposal. 
Gaussian proposals (such as drift changes) are severely inefficient for
multi-modal payouts (Capriotti 2008). The optimal proposal and
an LBFP estimate generated in the NPIS algorithm are shown in Figure
\ref{fig:StraddleOptProp}. The LBFP estimate closely approximates the optimal
proposal.
To account for the bimodality, we used $2h^{\text{opt}}$ as bin width in the 
QNPIS algorithm. However, $3h^{\text{opt}}$ gives only slightly worse results.
The simulation results for the strikes $K=$ 100 and $K=$ 110 are reported in
Table \ref{table:straddle}.
First notice, that NPIS significantly outperforms LSIS because of 
the better approximation of the optimal proposal. Second, the VR factors for
NPIS increase with sample size which agrees with Theorem 1.
Third, the combination of NPIS and QMC leads to massive efficiency gains.
Even after adjusting for the execution times the gains are enormous (see values for
the RCE).
Note, the increasing VR factors for QLSIS and QNPIS are a result of the QMC
sampling.

\begin{table}
\centering
\footnotesize
\begin{tabular}{ccccccc}
\toprule
\multicolumn{2}{c}{Parameters} & \multicolumn{5}{c}{VR (RCE)}\\ 
\cmidrule(r){1-2}
\cmidrule(r){3-7}
$N$ & $K$ & QMC & LSIS & NPIS & QLSIS & QNPIS \\
\hline
$2^{10}$ & 100 & 224 (380) & 1.3 (0.4) & 9 (0.9) & 1,064 (127) & 2.3 $\times
10^5$ (8,469) \\
		 & 110 & 253 (548) & 1   (0.4) & 6 (0.8) & 964   (150) & 3.2 $\times 10^5$
		 (1.5 $\times 10^4$) \\
         \hline
$2^{11}$ & 100 & 264 (557) & 1.3 (0.5) & 13 (1.7) & 1,361 (384) & 2.6 $\times
10^5$ (2.1 $\times 10^4$) \\ 
		 & 110 & 290 (532) & 1   (0.3) & 8  (1)   & 1,092 (291) & 3.1 $\times 10^5$
		 (2.4 $\times 10^4$)\\
         \hline
$2^{12}$ & 100 & 460 (941) & 1.3 (0.4) & 17 (2.2) & 2,209 (1,006) & 6.8
$\times 10^5$ (8.6 $\times 10^4$) \\ 
		 & 110 & 505 (953) & 1   (0.3) & 11 (1.3) & 2,201 (965) & 7.4 $\times 10^5$
		 (8.7 $\times 10^4$) \\
\bottomrule
\end{tabular}
\caption{\footnotesize The table reports the variance reduction (VR) factors and
the relative computational efficiency (RCE) for a straddle option within the Black-Scholes
model. Model parameters: $S(0) =$ 100, $\sigma = 0.3$, $r=0.05$,
$T=1$, and $|u|=d=1$. All values are computed based on 1,000 independent runs.}
\label{table:straddle}
\end{table}

\vspace{0.5cm}
\begin{center}
Example 2. Asian Options
\end{center}

An arithmetic Asian call with payout function
\[C_K(S) = \left(\frac1d \sum_{i=1}^{d} S(t_i) - K\right)^{+}\]
is investigated. The optimal proposal is unimodal. This integration problem is
well suited for NPIS/QNPIS because its ED is one.
The strikes $K=$ 100, 130, and 175 are
considered. For strike $K=$ 175 the option price is approximately 0.018 (for
$d=$16) representing a rare event option pricing framework (which is still of
practical interest). 

\begin{figure}
\centering
\includegraphics[height=200pt, keepaspectratio]{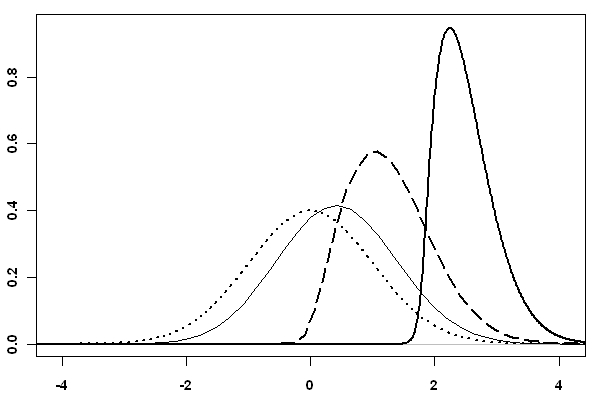}
\caption{\footnotesize Standard normal distribution (dotted line),
marginalized optimal proposal (of first principal component) for an Asian option
with strike $K=$ 60 (solid line), $K=$ 100 (dashed line), and $K=$ 140 (heavy line).
Model parameters: $S(0) =$ 100, $\sigma = 0.3$, $r=0.05$, $T=1$, and $d=16$.}
\label{fig:AsianOptProp}
\end{figure}

Table \ref{table:asian} shows the results for $d=$ 16 and
$d=$ 64 discretization steps. The results of the Gaussian IS algorithm (GIS)
based on saddle point approximation (Glasserman, Heidelberger, and Shahabuddin
1999) are also reported.
We emphasize that the VR and RCE increase with both strike level $K$ and the
sample size. The VR factors of GIS and LSIS are roughly constant. This coincides
with the theoretical results. Particularly in the rare event cases, massive
efficiency gains are achieved and NPIS/QNPIS improve significantly
over their parametric competitors. 
In addition, the values for RCE establish that NPIS and QNPIS are
computationally much more efficient than parametric IS strategies.
In the table, missing values indicate that the trial
stage sometimes failed to generate paths with positive payouts. 
To explain the result's dependency
on the strike level, the marginalized optimal proposal (of the first PCA
component) for different strikes were plotted (Figure \ref{fig:AsianOptProp}).
One can observe that both the mean and the variance of the marginalized optimal
proposals alter with $K$. As a result of the shrinking variance (and the
increasing skewness) of the marginalized optimal proposals, IS approaches based
on pure drift changes become worse (relatively to NPIS/QNPIS) as $K$ increases.

Table \ref{table:asianEqualTime} gives results for the case when the
execution time is fixed such as in real-time application. The
sample sizes were chosen so that all algorithms needed approximately the same
time for execution. The values suggest that the variance of NPIS is roughly ten times smaller
than those of existing IS techniques.

In Table \ref{table:asianKO}, the values for an Asian option with a knock-out
feature are shown. The option will pay nothing if the arithmetic average exceeds
the knock-out level $\tilde{K}$. The payout function is given by 
\[C_K(S) = \left(\frac1d \sum_{i=1}^{d} S(t_i) - K\right)^{+}
\mathbf{1}_{\{\frac1d \sum_{i=1}^{d} S(t_i) < \tilde{K} \}}.\]
The evaluation of this option is a difficult task because the relevant domain
is very narrow. The strike $K=$ 140 and the knock-out levels $\tilde{K}=$ 150
and $\tilde{K}=$ 170 are considered. The EDs are two and one for $\tilde{K}=$
150 and $\tilde{K}=$ 170, respectively. Both LSIS and NPIS have problems to
generate paths with positive payouts in the trial stage (which is reflected
in the missing values in Table \ref{table:asianKO}). Again, QNPIS significantly
improves over QLSIS.

Finally, simulations for an Asian straddle option that pays
\[
C_K(S) = (\frac1d \sum_{i=1}^{d} S(t_i)-K)^{+} + (K-\frac1d \sum_{i=1}^{d} S(t_i))^{+}
\]
are discussed. As for the standard straddle option, NPIS provides efficiency
gains compared with LSIS (see Table \ref{table:asianStraddle}). Although, the
VR factors and the RCE of QNPIS are large, they are much smaller than those
obtained for the standard straddle option. 

\begin{table}
\centering
\scriptsize
\begin{tabular}{cccccccccc}
\toprule
\multicolumn{3}{c}{Parameters} & & \multicolumn{6}{c}{VR
(RCE)}\\
\cmidrule(r){1-3}
\cmidrule(r){5-10}
$N$ & $d$ & $K$ & ED & QMC & GIS & LSIS & NPIS & QLSIS & QNPIS \\
\hline
$2^{10}$ & 16 & 100 & 1 & 139 (139) & 10 (3) & 9 (2) & 21 (11) & 1,427 (113) &
859 (187) \\ 
            & & 140 & 1 & 17 (17) & 55 (17) & 50 (10) & 200 (102) & 4,778 (375)
            & 5,462 (1,193) \\ 
            & & 175 & 1 & 2 (2) & 683 (202) & - (-) & 3,809 (1,941) & 4.3
            $\times 10^4$ (3,326) & 1.1 $\times 10^5$ (2.5 $\times 10^4$) \\ 
         & 64 & 100 & 1 & 145 (144) & 8 (3) & 8 (2) & 20 (11) & 1,409 (108) &
         909 (224) \\ 
            & & 140 & 1 & 16 (16) & 61 (19) & 53 (10) & 245 (138) & 5,679 (434)
            & 7,428 (1,828)
            \\ 
            & & 175 & 1 & 2 (2) & 902 (280) & - (-) & 4,403 (2,501) & 5.8
            $\times 10^4$ (4,506) & 1.0 $\times 10^5$ (2.5 $\times 10^4$) \\
            \hline
$2^{11}$ & 16 & 100 & 1 & 171 (173) & 9 (3) & 9 (2) & 28 (14) & 1,535 (226) &
908 (322) \\ 
            & & 140 & 1 & 21 (22) & 57 (17) & 52 (10) & 285 (146) & 5,647 (823)
            & 6,443 (2,267) \\ 
            & & 175 & 1 & 3 (3) & 680 (204) & - (-) & 5,161 (2,646) & 4.5
            $\times 10^4$ (6,599) & 1.3 $\times 10^5$ (4.4 $\times 10^4$) \\ 
         & 64 & 100 & 1 & 185 (185) & 9 (3) & 9 (2) & 30 (17) & 1,583 (225) &
         912 (360) \\ 
            & & 140 & 1 & 21 (16) & 69 (21) & 55 (10) & 329 (185) & 5,951 (847)
            & 8,027 (3,164) \\ 
            & & 175 & 1 & 2 (2) & 1,072 (332) & - (-) & 7,255 (4,117) & 6.2
            $\times 10^4$ (8,757) & 1.1 $\times 10^5$ (4.4 $\times 10^4$) \\
            \hline
$2^{12}$ & 16 & 100 & 1 & 339 (339) & 9 (3) & 9 (2) & 33 (17) & 2,549 (647) &
1,499 (767) \\ 
            & & 140 & 1 & 42 (43) & 56 (17) & 59 (12) & 324 (167) & 8,742
            (2,219) & 1.0 $\times 10^4$ (5,212) \\
            & & 175 & 1 & 5 (5) & 756 (232) & - (-) & 5,224 (2,696) & 8.7
            $\times 10^4$ (2.2 $\times 10^4$) & 2.2 $\times 10^5$ (1.1 $\times
            10^5$) \\ 
         & 64 & 100 & 1 & 354 (352) & 10 (3) & 10 (2) & 35 (20) & 2,743 (682) &
         1,627 (921) \\ 
            & & 140 & 1 & 36 (36) & 68 (21) & 57 (11) & 369 (209) & 9,685
            (2,407) & 1.3 $\times 10^4$ (7,388) \\ 
            & & 175 & 1 & 4 (4) & 1,031 (318) & - (-) & 7,414 (4,198) & 9.7
            $\times 10^4$ (2.4 $\times 10^4$) & 1.8 $\times 10^5$ (1.0 $\times
            10^5$) \\
\bottomrule
\end{tabular}
\caption{\footnotesize The table reports the variance reduction (VR) factors,
the relative computational efficiency (RCE), and the effective dimension (ED) for an Asian option within
the Black-Scholes model.
Model parameters: $S(0) =$ 100, $\sigma = 0.3$, $r=0.05$, $T=1$.
All values are computed based on 1,000 independent runs.}
\label{table:asian}
\end{table}

\begin{table}
\centering
\footnotesize
\begin{tabular}{cccccc}
\toprule
& & \multicolumn{4}{c}{VR ($N$)}\\
\cmidrule(r){3-6}
Time & ED & MC & GIS & LSIS & NPIS \\
\hline
0.35 & 1 & 1 ($2^{13}$) & 16 ($\lfloor 2^{11.19} \rfloor$) & 12 ($\lfloor
2^{10.68} \rfloor$) & 168 ($2^{12}$) \\ 
0.7  & 1 & 1 ($2^{14}$) & 16 ($\lfloor 2^{12.19} \rfloor$) & 11 ($\lfloor
2^{11.68} \rfloor$) & 175 ($2^{13}$) \\ 
1.4  & 1 & 1 ($2^{15}$) & 17 ($\lfloor 2^{13.19} \rfloor$) & 11 ($\lfloor
2^{12.68} \rfloor$) & 158 ($2^{14}$) \\
\bottomrule
\end{tabular}
\caption{\footnotesize The table reports the variance reduction (VR) factors,
the sample sizes ($N$), and the effective dimension (ED) for an Asian option
within the Black-Scholes model. The execution time is fixed to 0.35, 0.7, and 1.4 seconds, respectively. The
sample sizes are chosen such that all algorithms approximately achieved the
fixed execution time. Model parameters: $S(0) =$ 100, $\sigma = 0.3$, $r=0.05$,
$T=1$, $K = $ 140, and $d=$ 16. All values are computed based on 1,000
independent runs.}
\label{table:asianEqualTime}
\end{table}

\begin{table}
\centering
\footnotesize
\begin{tabular}{cccccccc}
\toprule
\multicolumn{2}{c}{Parameters} & & \multicolumn{5}{c}{VR
(RCE)}\\
\cmidrule(r){1-2}
\cmidrule(r){4-8}
$N$ & $\tilde{K}$ & ED & QMC & LSIS & NPIS & QLSIS & QNPIS \\
\hline
$2^{10}$ & 150 & 2 & 5 (5) & - (-) & - (-) & 69 (5) & 110 (21) \\
         & 170 & 1 & 16 (16) & - (-) & 37 (19) & 1,003 (80) & 1,362 (297) \\ 
         \hline
$2^{11}$ & 150 & 2 & 6 (6) & - (-) & - (-) & 68 (10) & 123 (36) \\
         & 170 & 1 & 18 (18) & - (-) & 134 (68) & 1,168 (171) & 1,613 (568) \\ 
         \hline
$2^{12}$ & 150 & 2 & 6 (6) & - (-) & - (-) & 82 (21) & 163 (63) \\
         & 170 & 1 & 23 (24) & - (-) & 106 (55) & 1,530 (394) & 1,883 (961) \\ 
\bottomrule
\end{tabular}
\caption{\footnotesize The table reports the variance reduction (VR) factors,
the relative computational efficiency (RCE), and the effective dimension (ED) for an Asian option
with a knock-out feature within the Black-Scholes model. Model parameters:
$S(0) =$ 100, $\sigma = 0.3$, $r=0.05$, $T=1$, $K = $ 140, and $d=$ 16. All
values are computed based on 1,000 independent runs.}
\label{table:asianKO}
\end{table}

\begin{table}
\centering
\footnotesize
\begin{tabular}{cccccccc}
\toprule
\multicolumn{2}{c}{Parameters} & & \multicolumn{5}{c}{VR
(RCE)}\\
\cmidrule(r){1-2}
\cmidrule(r){4-8}
$N$ & $d$ & ED & QMC & LSIS & NPIS & QLSIS & QNPIS \\
\hline
$2^{10}$ & 16 & 1 & 193 (199) & 1.2 (0.2) & 6 (3) & 300 (24) & 323 (71) \\
         & 64 & 1 & 213 (214) & 1.1 (0.2) & 6 (4) & 321 (25) & 361 (90) \\ 
         \hline
$2^{11}$ & 16 & 1 & 225 (233) & 1.2 (0.2) & 8 (4) & 359 (53) & 418 (151) \\
         & 64 & 1 & 256 (249) & 1.2 (0.2) & 9 (5) & 397 (57) & 410 (164) \\ 
         \hline
$2^{12}$ & 16 & 1 & 425 (440) & 1.2 (0.2) & 10 (5) & 634 (165) & 711 (372) \\
         & 64 & 1 & 454 (455) & 1.2 (0.2) & 11 (6) & 715 (179) & 717 (406) \\ 
\bottomrule
\end{tabular}
\caption{\footnotesize The table reports the variance reduction (VR) factors,
the relative computational efficiency (RCE), and the estimated effective
dimension (ED) for an Asian straddle option within the Black-Scholes model.
Model parameters: $S(0) =$ 100, $\sigma = 0.3$, $r=0.05$, $T=1$, and $K = $ 100.
All values are computed based on 1,000 independent runs.}
\label{table:asianStraddle}
\end{table}

\vspace{0.5cm}
\begin{center}
Example 3. Multi-Asset Options
\end{center}

In this example, multi-asset options are considered. 
Suppose one deals with $s$ assets that satisfy
\[ S_i(t) = S_i(0) \exp[(r-0.5\sigma^2)t + \sigma \sqrt{t} Z_i] \qquad
i=1,\ldots,s,\] 
where the correlation matrix of $Z_1,\ldots, Z_s$ is denoted by $\Sigma$.
To keep the setting simple, $S_i(0)=$ 100 and $\text{corr}(Z_i, Z_j)= 0.3$ for
$i,j=1,\ldots,s$, $i\not= j$ is assumed. The ED is reduced by applying PCA to
the correlation matrix. 
We investigate two different payout structures.
First, the price for an average option with payout
\[C_K(S_1, \ldots, S_s) = \left(\frac1s \sum_{i=1}^{s} S_i(T) - K\right)^{+}\]
is computed. 
The second option depends on the maximum of the underlyings' final values and
has the payout function \[C_K(S_1, \ldots, S_s) = \left(\max_{1\leq i \leq s}\{S_i(T)\} -
K\right)^{+}.\] From Table \ref{table:multiassetmean}, one can observe that
the results for the average option are qualitatively similar to those of the Asian option in
Example 2. Particularly, the ED is also equal to one.

\begin{figure}
\centering
\includegraphics[height=200pt, keepaspectratio]{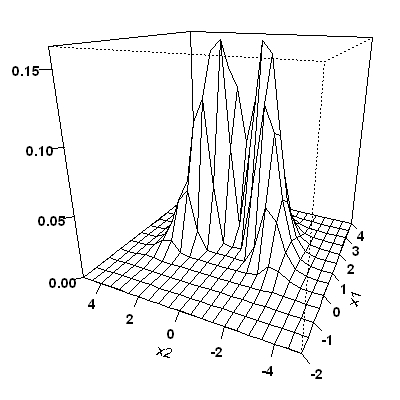}
\caption{\footnotesize LBFP estimate of the optimal proposal
for the multi-asset max option with strike $K=$ 150. Model parameters:  $S_i(0)
=$ 100 ($i=1,2$), $\sigma = 0.3$, $r=0.05$, $T=1$, and $|u|=2$.}
\label{fig:MultiAssetOptProp}
\end{figure}

The results for the second option with strikes $K=$ 150
and $K=$ 200 are reported in Tables \ref{table:multiassetmax150} and
\ref{table:multiassetmax200}, respectively. 
The pricing of the second option is a difficult problem because the ED is equal
to the nominal dimension.
Although, for $K=$ 200 QNPIS is superior to QMC and QLSIS for $s=$ 2, 3, and 4
(in terms of the VR factors), for $K=$ 150 this only holds for $s=$ 2 and 3. 
We emphasize on the massive efficiency gains obtained by QNPIS for strike $K=$
200. For $s>2$ the sample size used was too small for NPIS to perform well. 
We conclude that the applicability of NPIS/QNPIS depends not only on the ED
of the problem but also on the sample size used. 
An LBFP estimate of the optimal proposal for the case $s=2$ is plotted in
Figure \ref{fig:MultiAssetOptProp}. Here, the PCA construction leads to a
bimodal optimal proposal which can be closely approximated though an LBFP.

\begin{table}
\centering
\footnotesize
\begin{tabular}{cccccccc}
\toprule
\multicolumn{2}{c}{Parameters} & & \multicolumn{5}{c}{VR
(RCE)}\\
\cmidrule(r){1-2}
\cmidrule(r){4-8}
$N$ & $K$ & ED & QMC & LSIS & NPIS & QLSIS & QNPIS \\
\hline
$2^{10}$ & 100 & 1 & 179 (346) & 9 (4) & 24 (24) & 4,048 (620) & 3,315 (1,384)\\
         & 140 & 1 & 19 (38) & 43 (16) & 212 (210) & 5,171 (788) & 6,269
         (2,612)\\ 
         & 175 & 1 & 2 (3) & - (-) & 3,277 (3,229) & 2.5 $\times 10^4$ (3,856) &
         4.5 $\times 10^4$ (1.9 $\times 10^4$)\\
           \hline
$2^{11}$ & 100 & 1 & 212 (409) & 9 (3) & 34 (33) & 4,249 (1,197) & 3,677
(2,475)\\ 
         & 140 & 1 & 26 (50) & 48 (18) & 338 (333) & 5,438 (1,533) & 6,932
         (4,697)\\ 
         & 175 & 1 & 2 (4) & - (-) & 4,637 (4,571) & 2.9 $\times 10^4$ (8,313) &
         4.9 $\times 10^4$ (3.3 $\times 10^4$)\\
         \hline
$2^{12}$ & 100 & 1 & 428 (830) & 9 (3) & 49 (48) & 4,872 (2,403) & 3,996
(3,948)\\ 
         & 140 & 1 & 49 (96) & 52 (20) & 372 (368) & 6,373 (3,157) & 7,720
         (7,630)\\ 
         & 175 & 1 & 4 (9) & - (-) & 5,953 (5,857) & 4.3 $\times 10^4$ (2.1
         $\times 10^4$) & 6.5 $\times 10^4$ (6.4 $\times 10^4$)\\
\bottomrule
\end{tabular}
\caption{\footnotesize The table reports the variance reduction (VR) factors,
the relative computational efficiency (RCE), and the estimated effective
dimension (ED) for a multi-asset average option within the Black-Scholes model.
Model parameters: $S_i(0) =$ 100 ($i=1,\ldots,s$), $\sigma = 0.3$, $r=0.05$,
$T=1$, and $s=$ 16. All values are computed based on 1,000 independent runs.}
\label{table:multiassetmean}
\end{table}

\begin{table}
\centering
\footnotesize
\begin{tabular}{cccccccc}
\toprule
\multicolumn{2}{c}{Parameters} & & \multicolumn{5}{c}{VR
(RCE)}\\
\cmidrule(r){1-2}
\cmidrule(r){4-8}
$N$ & $s$ & ED & QMC & LSIS & NPIS & QLSIS & QNPIS \\
\hline
 $2^{11}$ & 2 & 2 & 44 (68) & 6 (1.8) & 10  (0.6)    & 145 (33) & 3,070
 (228)\\ 
          & 3 & 3 & 26 (56) & 3 (1.2) & 0.3 (0.02)   & 52 (14)  & 72 (7) \\
          & 4 & 4 & 24 (41) & 3 (1)     & 0.03 (0.002) & 28 (7)   & 7 (0.5) \\
          \hline
 $2^{12}$ & 2 & 2 & 76  (136) & 5 (1.9) & 25 (1.6)     & 213 (91)  & 5,848
 (620)\\ 
          & 3 & 3 & 42 (79)  & 3 (1.2) & 0.3 (0.02)   & 72 (32)  & 148 (16)
          \\ 
          & 4 & 4 & 27 (45) & 3 (1.1)   & 0.05 (0.002) & 36 (16)  & 8 (0.7) \\
          \hline
 $2^{13}$ & 2 & 2 & 211 (391) & 6 (2) & 61 (4)   & 396 (270)  & 4.7
 $\times 10^4$ (5,916) \\ 
          & 3 & 3 & 70 (119)  & 3 (1.1) & 2 (0.08) & 95 (65) 
 & 161 (20) \\ 
          & 4 & 4 & 35 (60) & 3 (1) & 0.01 (0.001)  & 42 (30) & 7 (0.7)\\
\bottomrule
\end{tabular}
\caption{\footnotesize The table reports the variance reduction (VR) factors,
the relative computational efficiency (RCE), and the estimated effective dimension
(ED) for a multi-asset max option with strike $K=$ 150. 
Model parameters: $S_i(0) =$ 100 ($i=1,\ldots,s$), $\sigma = 0.3$, $r=0.05$,
and $T=1$. 
All values are computed based on 1,000 independent runs.}
\label{table:multiassetmax150}
\end{table}

\begin{table}
\centering
\footnotesize
\begin{tabular}{cccccccc}
\toprule
\multicolumn{2}{c}{Parameters} & & \multicolumn{5}{c}{VR
(RCE)}\\
\cmidrule(r){1-2}
\cmidrule(r){4-8}
$N$ & $s$ & ED & QMC & LSIS & NPIS & QLSIS & QNPIS \\
\hline
 $2^{11}$ & 2 & 2 & 8 (14) & - (-)   & 49  (3) & 65 (17) & 7,997
 (652)\\ 
          & 3 & 3 & 4 (8) & 2 (0.6)   & 0.2 (0.01) & 17 (4)  & 165 (14) \\
          & 4 & 4 & 5 (8) & 0.8 (0.3) & 0.1 (0.006) & 9 (2)   & 20 (1.5) \\
          \hline
 $2^{12}$ & 2 & 2 & 13 (23) & - (-) & 163 (10)     & 82 (33)  & 1.8 $\times
 10^4$ (1,837)\\ 
          & 3 & 3 & 7 (13) & 4 (1.5)    & 3 (0.2) & 22 (10)  & 259 (28)
          \\ 
          & 4 & 4 & 6 (9) & 4 (1.3)  & 0.2 (0.01) & 11 (4.7)  & 24 (2.2) \\
          \hline
 $2^{13}$ & 2 & 2 & 32 (58) & - (-)   & 304 (18)   & 95 (65)  & 1.1 $\times
 10^5$ (1.4 $\times 10^4$) \\ 
          & 3 & 3 & 11 (19) & 5 (1.7)   & 1.2 (0.06)   & 28 (19) 
 & 292 (36) \\ 
          & 4 & 4 & 8 (13)   & 4 (1.5) & 0.05 (0.002) & 13 (9.1) & 27 (2.8)\\
\bottomrule
\end{tabular}
\caption{\footnotesize The table reports the variance reduction (VR) factors,
the relative computational efficiency (RCE), and the estimated effective dimension
(ED) for a multi-asset max option with strike $K=$ 200. 
Model parameters: $S_i(0) =$ 100 ($i=1,\ldots,s$), $\sigma = 0.3$, $r=0.05$,
and $T=1$. 
All values are computed based on 1,000 independent runs.}
\label{table:multiassetmax200}
\end{table}

\vspace{0.5cm}
\begin{center}
Example 4. Cap in the CIR model
\end{center}

Finally, we consider the CIR interest rate model (Cox, Ingersoll, and Ross
1985). Here, interest rate $r_t$ follows a square-root diffusion model
\begin{equation*}
d r_t = \kappa(\theta-r_t) dt + \sigma \sqrt{r_t} d W_t.
\end{equation*}
The first order Euler discretization yields
\begin{equation*}
r_{t_{k+1}} = r_{t_{k}} + \kappa(\theta-r_{t_{k}}) \Delta t + \sigma
\sqrt{r_{t_{k}}} Z_{t_k},
\end{equation*}
with $Z_{t_k} \sim \mathcal{N}(0,1)$ and $\Delta t = T/d$. 
The aim is to evaluate the price of an interest rate cap. It pays $(r_{t_{k}}
- K)^{+}$ at time $t_{k+1}$ ($k=0,\ldots,d-1$) subject to strike $K$. 
The discounted payout is given by
\begin{equation*}
\sum_{i=0}^{d-1} \exp[-\Delta t \sum_{j=0}^{i} r_{t_{k}}] (r_{t_{k}} - K)^{+}.
\end{equation*}
The parameter values used in the simulations are $d=16$, $r_0=0.07$,
$\theta=0.075$, $\kappa=0.2$, $\sigma=0.02$, $T=$ 1 and 2, 
$K = 0.06$, $0.07$, and $0.08$. The results are reported in Table
\ref{table:circapd16} and Table \ref{table:circapd64}. Again the ED is
equal to one, which explains the good performance of NPIS/QNPIS.
In particular, QNPIS strongly outperforms QLSIS for small strike levels.

\begin{table}
\centering
\footnotesize
\begin{tabular}{ccccccccc}
\toprule
\multicolumn{3}{c}{Parameters} & & \multicolumn{5}{c}{VR
(RCE)}\\
\cmidrule(r){1-3}
\cmidrule(r){5-9}
$N$ & $T$ & $K$ & ED & QMC & LSIS & NPIS & QLSIS & QNPIS \\
\hline
$2^{11}$ & 1 & .05 & 1 & 230 (231) & 2 (0.4) & 0.7 (0.4) & 396 (58) & 2,313
(814) \\ 
           & & .06 & 1 & 271 (280) & 3 (0.6) & 3 (1.4) & 798 (119) & 2,828
           (1,021) \\ 
           & & .07 & 1 & 233 (236) & 9 (1.8) & 12 (6) & 287 (42) & 256 (91) \\
           & & .08 & 1 & 9 (9) & 51 (10) & 36 (19) & 458 (67) & 219 (78) \\
         & 2 & .05 & 1 & 232 (235) & 3 (0.5) & 1.2 (0.6) & 486 (71) & 2,754
         (977) \\ 
           & & .06 & 1 & 297 (298) & 5 (1) & 4 (2) & 820 (120) & 2,555 (905) \\
           & & .07 & 1 & 240 (247) & 10 (1.9) & 13 (7) & 281 (42) & 288 (104) \\
           & & .08 & 1 & 25 (25) & 25 (5) & 11 (6) & 300 (44) & 157 (56) \\
           \hline
$2^{12}$ & 1 & .05 & 1 & 479 (489) & 2 (0.4) & 1.1 (0.6) & 820 (210) & 5,235
(2,717) \\ 
           & & .06 & 1 & 582 (588) & 3 (0.6) & 4 (2) & 1,621 (414) & 4,961
           (2,081) \\ 
           & & .07 & 1 & 415 (426) & 9 (1.8) & 13 (7) & 388 (100) & 332 (174) \\
           & & .08 & 1 & 15 (15) & 49 (10) & 43 (22) & 588 (151) & 283 (146) \\
         & 2 & .05 & 1 & 484 (492) & 2 (0.4) & 1.9 (1) & 1,007 (257) & 5,723
         (2,957) \\ 
           & & .06 & 1 & 626 (634) & 5 (0.9) & 6 (3) & 1,377 (352) & 4,182
           (2,143) \\ 
           & & .07 & 1 & 422 (433) & 9 (1.9) & 14 (7) & 375 (97) & 360 (188) \\
           & & .08 & 1 & 44 (45) & 26 (5) & 18 (9) & 374 (96) & 214 (111) \\
\bottomrule
\end{tabular}
\caption{\footnotesize The table reports the variance reduction (VR) factors,
the relative computational efficiency (RCE), and the estimated effective
dimension (ED) for a cap within the CIR model. Model parameters:
$r_0 = 0.07$ , $\theta = 0.075$, $\kappa = 0.2$, $\sigma = 0.02$, and $d = 16$. 
All values are computed based on 1,000 independent runs.}
\label{table:circapd16}
\end{table}

\begin{table}
\centering
\footnotesize
\begin{tabular}{ccccccccc}
\toprule
\multicolumn{3}{c}{Parameters} & & \multicolumn{5}{c}{VR
(RCE)}\\
\cmidrule(r){1-3}
\cmidrule(r){5-9}
$N$ & $T$ & $K$ & ED & QMC & LSIS & NPIS & QLSIS & QNPIS \\
\hline
$2^{11}$ & 1 & .05 & 1 & 238 (239) & 3 (0.5) & 0.9 (0.5) & 451 (65) & 2,924
(1,164) \\ 
           & & .06 & 1 & 284 (284) & 4 (0.7) & 3 (1.8) & 940 (135) & 4,225
           (1.677) \\ 
           & & .07 & 1 & 263 (263) & 10 (2) & 15 (8) & 414 (59) & 335 (133) \\ 
           & & .08 & 1 & 9 (9) & 48 (9) & 30 (17) & 536 (77) & 336 (133) \\ 
         & 2 & .05 & 1 & 240 (239) & 3 (0.5) & 1.4 (0.8) & 552 (79) & 3,760
         (1,483) \\ 
           & & .06 & 1 & 309 (308) & 7 (1.2) & 5 (3) & 1,013 (145) & 3,345
           (1,325) \\ 
           & & .07 & 1 & 270 (269) & 11 (2) & 15 (8) & 410 (58) & 402 (158) \\ 
           & & .08 & 1 & 28 (27) & 25 (5) & 21 (12) & 411 (59) & 162 (64) \\ 
           \hline
$2^{12}$ & 1 & .05 & 1 & 471 (472) & 2 (0.4) & 1.3 (0.7) & 870 (218) & 7,202
(4,101) \\ 
           & & .06 & 1 & 571 (571) & 3 (0.6) & 5 (3) & 1,808 (453) & 7,422
           (4,219) \\ 
           & & .07 & 1 & 491 (491) & 10 (2) & 16 (9) & 532 (133) & 475 (271) \\ 
           & & .08 & 1 & 17 (17) & 43 (8) & 34 (19) & 674 (168) & 346 (196) \\ 
         & 2 & .05 & 1 & 477 (474) & 2 (0.4) & 2 (1.2) & 1,074 (266) & 9,622
         (5,442) \\ 
           & & .06 & 1 & 627 (624) & 6 (1.1) & 7 (4) & 1,371 (342) & 5,875
           (3,328) \\ 
           & & .07 & 1 & 507 (503) & 11 (2) & 16 (9) & 522 (130) & 531 (299) \\ 
           & & .08 & 1 & 51 (51) & 24 (5) & 15 (9) & 470 (117) & 177 (100) \\ 
\bottomrule
\end{tabular}
\caption{\footnotesize The table reports the variance reduction (VR) factors,
the relative computational efficiency (RCE), and the estimated effective
dimension (ED) for a cap within the CIR model. Model parameters: $r_0 = 0.07$
, $\theta = 0.075$, $\kappa = 0.2$, $\sigma = 0.02$, and $d = 64$. 
All values are computed based on 1,000 independent runs.}
\label{table:circapd64}
\end{table}

\vspace{0.5cm}

\begin{center}
10. CONCLUSION
\end{center}

An NPIS algorithm was proposed that applies NIS to a carefully chosen subspace.
The MSE convergence properties were explored. They establish the
asymptotic optimality of the approach and suggest that it improves over
parametric IS \-- at least \-- asymptotically. In particular, NPIS is shown to
achieve increasing efficiency compared to crude MC and parametric IS.
Its usefulness for practical sample sizes was verified through well-known
option pricing scenarios. Large VR factors were obtained in
certain situations. It was shown that NPIS is advantageous over existing IS
methods for problems with low ED, which is often the case in finance.
Particularly, situations of rare event dependency or multi-modal optimal proposals are well suited for
NPIS. There, existing methods often fail.
The combination of NPIS and QMC resulted in enormous efficiency gains.
A detailed description of the implementation of the proposed algorithm was
provided.
It is emphasized that NPIS can be applied ``blindly''. No analytical
investigation of the payout function is required. In addition, being generally
applicable NPIS is not restricted to a specific kind of diffusion model or payout function.
It can be applied to other settings occurring in finance, such as
the estimation of option sensitivities or the evaluation of the value-at-risk.

\vspace{0.5cm}

\begin{samepage}

\begin{center}
APPENDIX A: PROOF OF THEOREM 1
\end{center}

\noindent
{\bf Prerequisites for Theorem \ref{thm:npis}.}
The following quantities are not required in practical application. However,
they are necessary for the proof of Theorem \ref{thm:npis}.
\end{samepage}
Let $A_M$ be an increasing sequence of compact sets defined by $A_M
= \{ \mathbf{x} \in \RR^{|u|} : \breve{q}^{\text{opt}}(\mathbf{x}) \geq c_M \}$,
where $c_M > 0$ and $c_M \rightarrow 0$ as $M$ goes to infinity. For any
function $g$, we denote the restriction of $g$ on $A_M$ by $g_M$. Furthermore,
the volume of $A_M$ is denoted by $V_M$.
The NPIS estimator $\hat{I}_{\varphi_M}^{\text{NPIS}}$
is obtained by substituting $\hat{q}^{\text{opt}}$ (in the algorithm) for
\begin{equation*}
\hat{q}_M^{\text{opt}}(\mathbf{x}_{u}) = 
		\begin{cases}
        \frac{\hat{f}_M(\mathbf{x}_{u}) + \delta_M}{\frac1M \sum_{j=1}^M
        \omega_M^j + V_M \delta_M} \quad \text{for} & \mathbf{x}_{u} \in A_M,\\
		\qquad 0 & \text{else}.
        \end{cases}
\end{equation*}

\medskip

\noindent
{\em Assumption 1:}
$\breve{q}^{\text{opt}}$ has three continuous and square integrable
derivatives on its support and it is bounded. In addition, $\int
(\nabla^2 \breve{q}^{\text{opt}})^4 (\breve{q}^{\text{opt}})^{-3} < \infty$ 
where $\nabla^2 \breve{q}^{\text{opt}} = \partial^2 \breve{q}^{\text{opt}}/ \partial
\mathbf{x}_1^2 + \ldots + \partial^2 \breve{q}^{\text{opt}}/ \partial \mathbf{x}_d^2$.

\medskip 

\noindent
{\em Assumption 2:}
Trial distribution $q_0$ is chosen such that $\mathbf{E} [\breve{q}^{\text{opt}}
q_0^{-1}]^4$ is finite on $\text{supp}(\breve{q}^{\text{opt}})$.

\medskip 

\noindent
{\em Assumption 3:}
Sample sizes $M,N \rightarrow \infty$, bin width $h$ satisfies $h
\rightarrow 0$ and $M h^{|u|} \rightarrow \infty$. Additionally, we have
$\delta_M > 0$, $V_M \delta_M = o(h^2)$ and $M^3 (V_M \delta_M)^4 \rightarrow
\infty$.

\medskip 
\noindent
{\em Assumption 4:}
$c_M$ is chosen such that $\frac{h^8 + (Mh^{|u|})^{-2}}{\delta_M
c_M^3} = o(\frac{h^4 + (Mh^{|u|})^{-1}}{c_M})$ and $\frac{h^4 +
(Mh^{|u|})^{-1}}{c_M} \rightarrow 0$.

\medskip 
\noindent
{\em Assumption 5:}
The sequence $c_M$ guaranties $(\int \breve{q}^{\text{opt}}
\mathbf{1}_{\{\breve{q}^{\text{opt}}  < c_M \}})^2 = o(M^{-1} h^4 + (M^2
h^{|u|})^{-1})$.

\bigskip

\noindent
{\bf Proof of Theorem \ref{thm:npis}.}
Conditional on the samples $\{\tilde{\mathbf{x}}^1, \tilde{\mathbf{x}}^2,
\ldots, \tilde{\mathbf{x}}^M \}$, the variance of
$\hat{I}_{\varphi_M}^{\text{NPIS}}$ can be written as
\begin{eqnarray*}
\frac{\sigma^2_{\text{IS}}} {N}
&=& \frac{I_{\varphi}^2}{N} \int \left\{ \frac{\varphi(\mathbf{x})
p(\mathbf{x}_{u})}{I_{\varphi}} - \hat{q}_M^{\text{opt}}(\mathbf{x}_{u})
\right\}^2 \frac{p(\mathbf{x}_{-u})} {\hat{q}_M^{\text{opt}}(\mathbf{x}_{u})} d
\mathbf{x}\\ &=& \frac{I_{\varphi}^2}{N} \int \left[
\frac{\nu(\mathbf{x})^2}{I_{\varphi}^{2}}  + \left\{
\breve{q}^{\text{opt}}(\mathbf{x}_{u}) - 
\hat{q}_M^{\text{opt}}(\mathbf{x}_{u}) \right\}^2 \right]
\frac{p(\mathbf{x}_{-u})}  {\hat{q}_M^{\text{opt}}(\mathbf{x}_{u})} d \mathbf{x}
\end{eqnarray*}
with $\nu(\mathbf{x}) = \varphi(\mathbf{x}) p(\mathbf{x}_{u}) - \int
\varphi(\mathbf{x}) p(\mathbf{x}) d \mathbf{x}_{-u}$. 
The right term in the brackets (quantifying the nonparametric estimation
error) can be treated analogous to the proof of Theorem 2 in
Neddermeyer (2009). However, as a result of the integration with
respect to $\mathbf{x}_{-u}$, a different variance term is obtained. The optimal
bin width is derived through differentiation.

\vspace{0.5cm}

\begin{center}
APPENDIX B: DETERMINATION OF THE EFFECTIVE DIMENSION
\end{center}

An MC procedure to determine the ED is discussed (Wang and Fang 2003). It can
be shown that the cumulated variances satisfy
\[
\Gamma_u = \int_{\RR^{2d-|u|}} \varphi(\mathbf{x}) \varphi(\mathbf{x}_{u},
\mathbf{y}_{-u}) p(\mathbf{x}) p(\mathbf{y}) d \mathbf{x} d
\mathbf{y}_{-u} - I_\varphi^2,
\]
where both $\mathbf{x}$ and $\mathbf{y}$ are vectors in $\RR^d$.
Hence, the ED can be computed based on the approximations 
\[
\hat{\Gamma}_u = \frac{1}{l} \sum_{i=1}^{l} \varphi(\mathbf{x}^i)
\varphi(\mathbf{x}^i_{u}, \mathbf{y}^i_{-u}) - \hat{I}_\varphi^2
\qquad (i=1,2,\ldots,d)
\]
and 
\[
\hat{\sigma}^2(\varphi) = \frac{1}{l} \sum_{i=1}^{l} \varphi(\mathbf{x}^i)^2 -
\hat{I}_\varphi^2
\]
with $\hat{I}_\varphi = 1/l \sum \varphi(\mathbf{x}^i)$. The samples
$\mathbf{x}^1$, $\mathbf{x}^2$, \ldots, $\mathbf{x}^l$, $\mathbf{y}^1$,
$\mathbf{y}^2$, \ldots, $\mathbf{y}^l$ are drawn from $p$.

\newpage

\begin{center}
REFERENCES
\end{center}

\begin{description}

\item Adler, R.\ J. (1990), ``An Introduction to Continuity, Extrema, and
Related Topics for General Gaussian Processes,'' Hayward, California: Institute
of Mathematical Statistics.

\item Avramidis, A.\ N. (2002), ``Importance Sampling for Multimodal Functions
and Application to Pricing Exotic Options,'' in {\it Winter Simulation Conference Proceedings}, 1493-1501.

\item Caflisch, R.\ E., Morokoff, W.\ J., and Owen, A.\ B. (1997),  ``Valuation
of mortgage backed securities using Brownian bridges to reduce effective
dimension,'' in {\it Journal of Computational Finance}, 1, 27-46.

\item Capriotti, L. (2008), ``Least Squares Importance Sampling for Monte
Carlo Security Pricing,'' in {\it Quantitative Finance}, 8, 485-497.

\item Cox, J., Ingersoll, J.\ E., and Ross, S.\ A. (1985),  ``A Theory of the
Term Structure of Interest Rates,'' in {\it Econometrica}, 53, 385-407.

\item Givens, G.\ H., and Raftery, A.\ E. (1996),  ``Local Adaptive Importance
Sampling for Multivariate Densities With Strong Nonlinear Relationships,'' in
{\it Journal of American Statistical Association}, 91, 132-141.

\item Glasserman, P. (2004), {\it
Monte Carlo Methods in Financial Engineering}, New York: Springer.

\item Glasserman, P., Heidelberger, P., and Shahabuddin, P. (1999),
``Asymptotically optimal importance sampling and stratification for pricing
path-dependent options,'' in {\it Mathematical Finance}, 9, 117-152.

\item J{\"a}ckel, P. (2002), {\it Monte Carlo methods in finance}, West Sussex:
Wiley.

\item Kim, Y.\ B., Roh, D.\ S., and Lee, M.\ Y. (2000), ``Nonparametric Adaptive
Importance Sampling For Rare Event Simulation,'' in {\it Winter Simulation
Conference Proceedings}, 767-772.

\item L'Ecuyer, P. (1994), ``Efficiency improvement and variance reduction,''
in {\it Winter Simulation Conference Proceedings}, 122-132.

\item Matou{\v{s}}ek, J. (1998), ``On the $L_2$-discrepancy for anchored
boxes,'' in {\it Journal of Complexity}, 14, 527-556.

\item Matsumoto, M., and Nishimura, T. (1998), ``Mersenne Twister: A
623-Dimensionally Equidistributed Uniform Pseudo-Random Number Generator,'' in
{\it ACM Transactions on Modeling and Computer Simulations}, 8, 3-30.

\item Moro, B. (1995), ``The full monte,'' in {\it Risk}, 8, 57-58.

\item Neddermeyer, J.\ C. (2009), ``Computationally Efficient Nonparametric
Importance Sampling,'' in {\it Journal of American Statistical Association}, in
press.

\item Niederreiter, H. (1992), {\it Random Number Generation and Quasi-Monte
Carlo Methods}, Philadelphia: Society for Industrial and Applied Mathematics.

\item Ninomiya, S., and Tezuka, S. (1996), ``Towards real time pricing of
complex financial derivatives,'' in {\it Applied Mathematical Finance}, 3, 1-20.

\item {\"O}kten, G., and Eastman, W. (2004), ``Randomized {Quasi-Monte Carlo}
methods in pricing securities,'' in {\it Journal of Economic Dynamics \&
Control}, 28, 2399-2426.

\item Owen, A.\ B. (1992), ``Orthogonal arrays for computer experiments,
integration and visualization,'' in {\it Statistica Sinica}, 2, 439-452.

\item --- (1995), ``Randomly Permuted (t; m; s)-Nets and (t; s)- Sequences,'' in
 {\it Monte Carlo and Quasi-Monte Carlo Methods in Scientific Computing}, eds.
 \ H. Niederreiter and P. J.-S. Shiue, New York: Springer, 299-317.

\item Paskov, S.\ H., and Traub, J.\ F. (1995), ``Faster valuation of financial
derivatives,'' in {\it Journal of Portfolio Management}, 22, 113-120.

\item Press, W.H., Teukolsky, S.A., Vetterling, W.T., and Flannery, B.P. (1992),
{\it Numerical Recipes in C (2nd ed.)}, Cambridge: Cambridge University Press.

\item Robert, C.\ P., and Casella, G. (2004), {\it
Monte Carlo Statistical Methods}, New York: Springer.

\item Rubinstein, R.\ Y. (1981), {\it
Simulation and the Monte Carlo Method}, New York: Wiley.

\item Sobol, I.\ M. (1967), ``On the distribution of points in a cube and the
approximate evaluation of integrals,'' in {\it USSR Journal of Computational Mathematics
and Mathematical Physics}, 7, 784-802.

\item Su, Y., and Fu, M.\ C. (2000), ``Importance sampling in derivative
securities pricing,'' in {\it Winter Simulation Conference Proceedings}, 587-596.

\item --- (2002), ``Optimal importance sampling in securities
pricing,'' in {\it Journal of Computational Finance}, 5, 27-50.

\item Takemura, A. (1993), ``Tensor analysis of ANOVA decomposition,'' in {\it Journal
of American Statistical Association}, 88, 1392-1397.

\item Terrell, G.R. (1983), ``The Multilinear Frequency Spline,'' Technical
Report, Rice University, Dept. of Math Sciences.

\item Traub, J.\ F., and Werschulz, A.\ G. (1998), {\it Complexity and
Information}, Cambridge: Cambridge University Press.

\item Vazquez-Abad, F., and Dufresne, D. (1998), ``Accelerated Simulation for
Pricing Asian Options,'' in {\it Winter Simulation Conference Proceedings}, 1493-1500.

\item Wang, X., and Fang, K.-T. (2003), ``The effective dimension and
quasi-Monte Carlo integration,'' in {\it Journal of Complexity}, 19, 101-124.

\item Zhang, P. (1996), ``Nonparametric Importance Sampling,'' in {\it Journal
of American Statistical Association}, 91, 1245-1253.

\end{description}

\end{document}